# Dynamical stability and electronic structure of *β*-phosphorus carbide nano-wires

Shcherbinin S. A.[1], Ustiuzhanina S.V.[2], Kistanov A. A.*,[3]
*1 – Southern Federal University, 344006 Rostov-on-Don, Russia*
*2 – Institute for Metals Superplasticity Problems, Russian Academy of Sciences, 450001 Ufa, Russia*
*3 – Nano and Molecular Systems Research Unit, University of Oulu, 90014 Oulu, Finland*
*andrey.kistanov@oulu.fi*

**Abstract** In this work, β-phosphorus carbide 1D nano-wires (PCNWs) are investigated in the framework of density functional theory. The dynamical stability of the considered β-PCNWs at 300 K is verified using *ab initio* molecular dynamics calculations. According to the results on the band structure calculations, β-PCNWs can be semiconductors, semimetals or metals depending on their size and form. Thus, owning to their unique shape and high tunability of electronic properties β-PCNWs may be used in optical and photovoltaic nanodevices.

## 1. Introduction

In the last decade, the number of new two-dimensional (2D) materials has been constantly growing. Advanced theoretical and novel experimental approaches make it possible to obtain hybrid 2D materials [1-4]. Recent theoretical predictions [5,6] and experimental investigations [7] have proven the existence of several allotropes of a new 2D material, phosphorus carbide (PC). Depending on its structure PC can be metallic, semi-metallic or semiconductor [5].

The predicted allotropes exhibit thermal stability and possess well tunable electronic structure [8]. In addition, the possibility of rolling of α-PC monolayer to a PC nanotube at room temperature under compressive strain has been found has been shown [9]. Another predicted *γ*-allotrope of PC with an InSe-like structure has been shown to have promoted adsorption of lithium atoms, which render its application in rechargeable batteries [6]. Furthermore, *γ*-PC has been found as a good material for gas sensing and storage devices [10]. Very recently, β-PC has been fabricated via a novel carbon doping technique [11]. Theoretical and experimental studies have shown the existence of different β-PC phases which may have a potential for application in nanodevices [11, 12].

In this work using first-principles calculations we studied the structural stability and electronic band structure of $\beta_0$- and $\beta_1$-PC 1D nano-wires.

## 2. Simulation Details

The computational simulations were performed by using the Vienna ab initio simulation package (VASP) [13] within the framework of the density functional theory.

one need to evince that it cannot undergo any structural changes that lower its energy. In practice, at the first-principles level, the verification can be done using AIMD simulations

To evince the dynamical stability of the considered structures, the ab initio molecular dynamics calculations [14] which is the most common method for low dimensional materials [15] were implemented. The structure optimization and band structure calculations the Perdew–Burke–Ernzerhof (PBE) functional with generalized gradient approximation (GGA) [16] was selected with an energy cutoff of 400 eV. All the structures were fully optimized until the forces become smaller than 0.01 eV/Å.

The optimized unicells of the considered $\beta_0$- and $\beta_1$ allotropes of PC are shown in Figure 1. The calculated lattice constants of $\beta_0$- and $\beta_1$-PC are *a* = 5.050 Å and *b* = 2.915 Å and *a* =

4.725 and $b = 2.915$, respectively. The results are in good agreement with the available references [4].

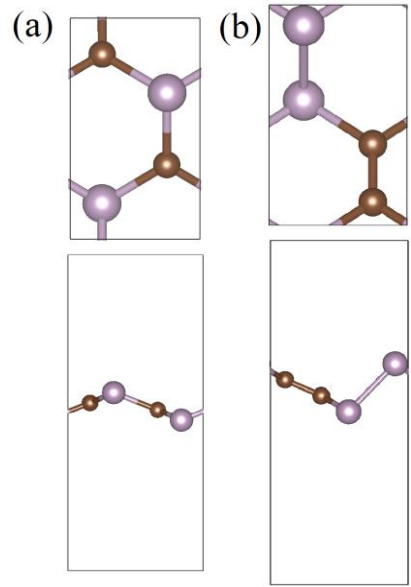

**Figure 1.** The optimized unicells of the considered (a) $\beta_0$- and (b) $\beta_1$ allotropes of PC.

## 3. Simulation Results

First, we created $\beta_0$- and $\beta_1$-PCNWs consisting of 12 atoms and more by the rippling of $\beta_0$-PC and $\beta_1$-PC monolayers along their armchair (APCNW) and zigzag (ZPCNW) directions. Further, the dynamical stability of the created PCNWs is systematically checked using AIMD calculations conducted at 300 K during the period time 10 ps. It is found that the stable $\beta_0$- and $\beta_1$-APCNWs of the smallest/biggest size consist of 12/40 atoms. Stable $\beta_0$-ZPCNWs may consist of 32 to 44 atoms, while $\beta_1$-ZPCNWs consists of 24 to 44 atoms. For PCNWs found to be stable, we next calculated the band structure. Figure 2 presents the bandgap size of the considered PCNWs as a function of their size.

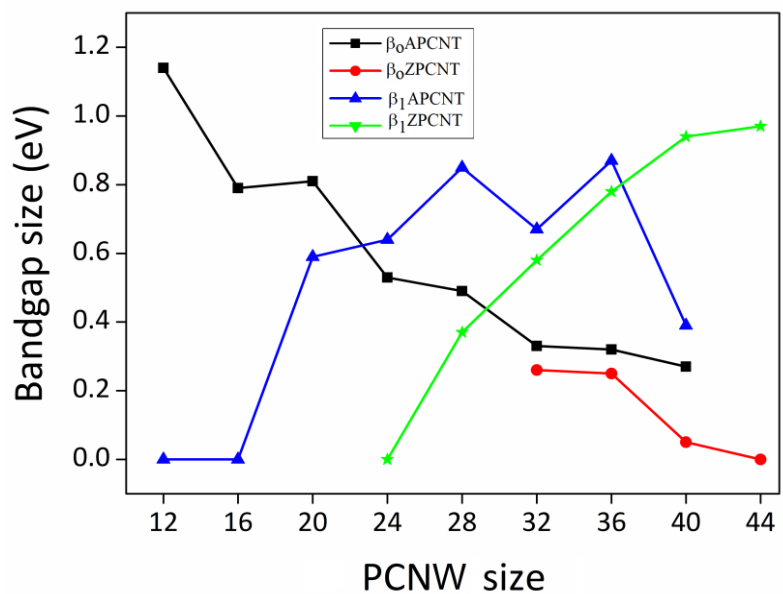

**Figure 2.** The bandgap size as a function of size of $\beta_0$- and $\beta_1$-APCNW and $\beta_0$- and $\beta_1$-ZPCNW.

Figure 3 shows the optimized atomic (upper panels) and band (lower panels) structures of $\beta_0$-APCNWs. As it is seen from Figure 3, $\beta_0$-APCNWs may have different forms such as triangle and star-like structure. Depending on the size, the $\beta_0$-APCNWs vary from a direct (wire consisting of 12 atoms) to an indirect bandgap semiconductor. According to Figure 2 with increasing the size of $\beta_0$-APCNWs from 12 atoms to 40 atoms its bandgap size decreases drastically from 1.14 eV to 0.27 eV.

The atomic and band structures of stable $\beta_1$-APCNWs are presented in Figure 4. Based on the geometry optimization results, the APCNWs are characterized by a star-like structure. The band structure calculations (Figure 2) suggest $\beta_1$-APCNWs may have a zero bandgap (wires consisting of 12 and 16 atoms), or to be a direct (wires consisting of 20 atoms) and an indirect (wires consisting of 24 to 40 atoms) bandgap semiconductors. Based on the results presented in Figure 2, the bandgap size of $\beta_1$-APCNWs is increasing by leaps from 0 eV up to 0.87 eV with increasing their size.

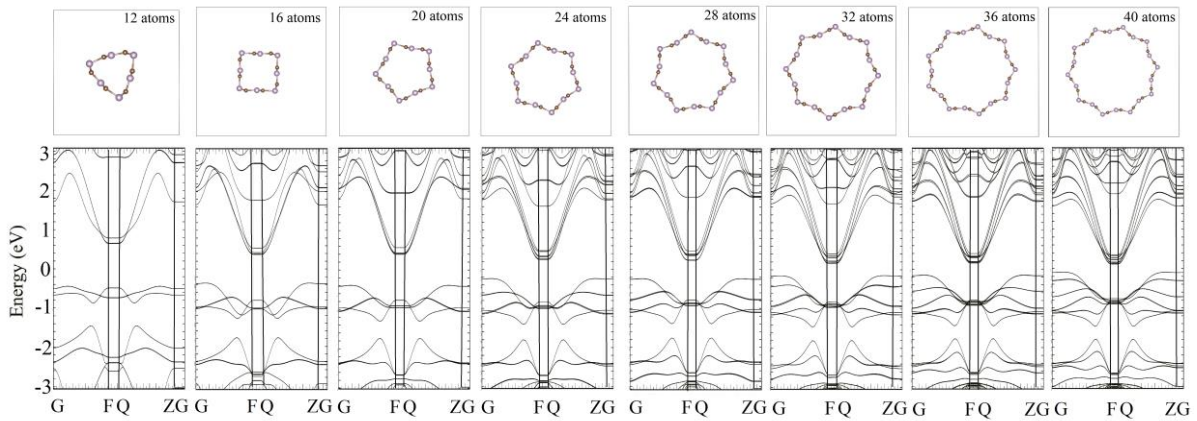

**Figure 3.** The atomic (the upper panel) and band (the lower panel) structures of $\beta_0$-APCNWs.

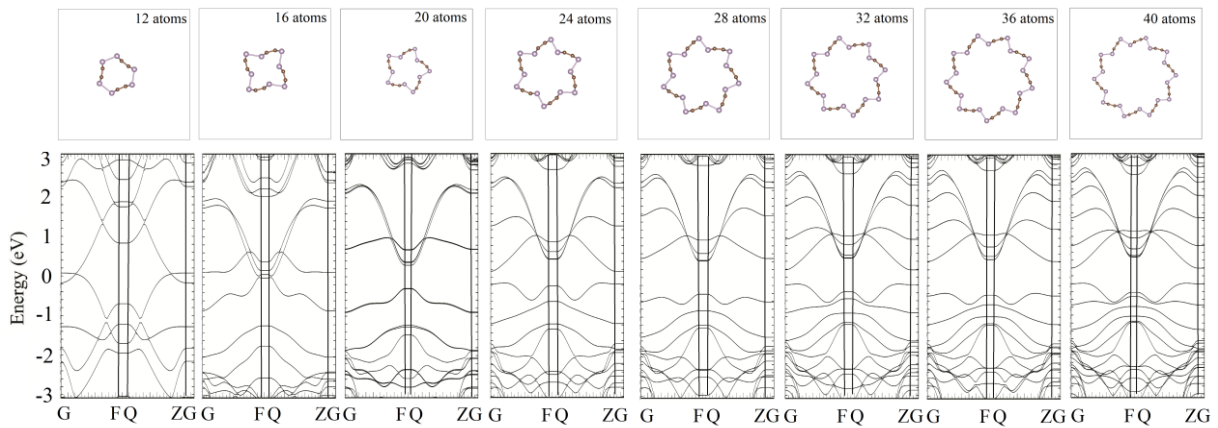

**Figure 4.** The atomic (the upper panel) and band (the lower panel) structures of $\beta_1$-APCNWs.

ZPCNWs have been found less stable then APCNWs. In case of $\beta_0$-ZPCNWs, there are only four stable configurations, which are shown in Figure 5. The $\beta_0$-ZPCNWs consisting of 32 to 40 atoms are direct bandgap semiconductors while $\beta_0$-ZPCNW consisting of 44 atoms has a zero bandgap (see Figure 5, lower panels). With increasing the size, the bandgap size of $\beta_0$-ZPCNWs decreases as it is shown in Figure 2. $\beta_0$-ZPCNWs have six stable configurations whish are presented in Figure 6. The most interesting results are found for band structure of $\beta_1$-ZPCNWs. It is predicted that $\beta_1$-ZPCNW consisting of 24 atoms is metallic. $\beta_1$-ZPCNWs consisting of 28 to 40 atoms are indirect bandgap semiconductors, and $\beta_1$-ZPCNW consisting of 44 atoms is a direct bandgap semiconductor (see Figure 6, lower panels). Differently from other PCNWs here the bandgap size of $\beta_1$-ZPCNWs significantly increases with size from 0 eV to 0.97 eV (see Figure 2).

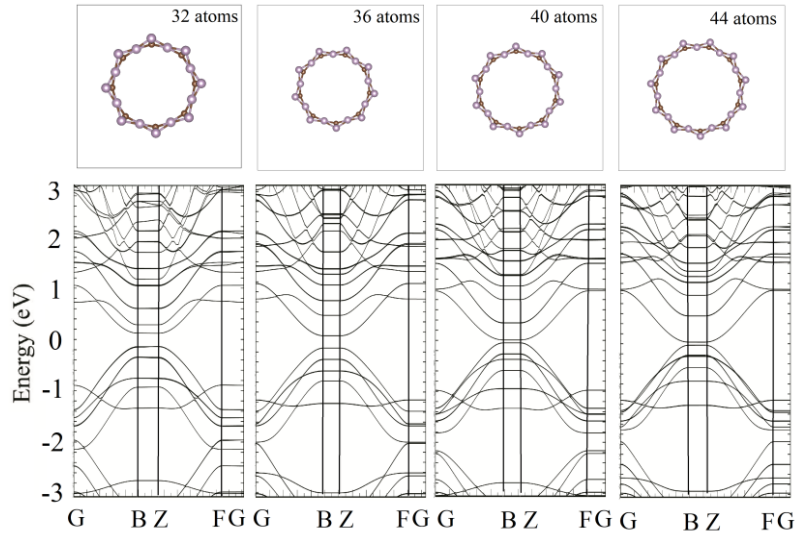

**Figure 5.** The atomic (the upper panel) and band (the lower panel) structures of $\beta_0$-ZPCNWs.

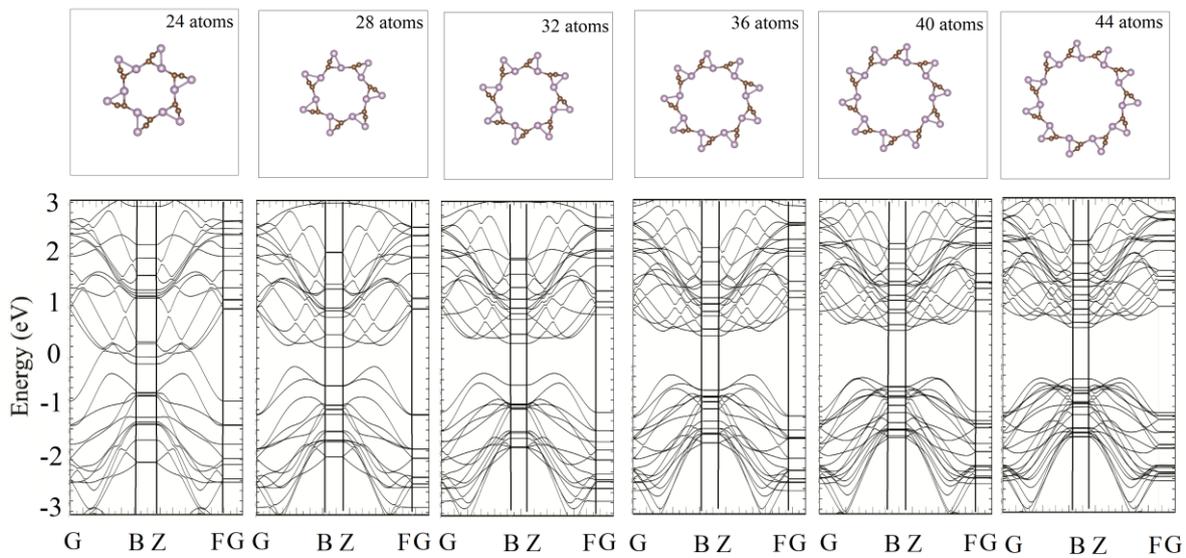

**Figure 6.** The atomic (the upper panel) and band (the lower panel) structures of $\beta_1$-ZPCNWs.

## 4. Conclusions

In conclusion, our theoretical predictions show the existence of β-PCNWs of different sizes and unique shapes. These β-PCNWs also possess existing electronic properties. Particularly, the bandgap size of β-PCNWs is directly depends on their size. Such tunability of the band structure suggests β-PCNWs as a perfect material for application in optoelectronic nanodevices.

**Acknowledgments**
For A.A.K. this work is supported by the Academy of Finland grant No. 311934. S.A.Sh. acknowledges the financial support by the Ministry of Education and Science of the Russian Federation (state task in the field of scientific activity, Southern Federal University), theme N BAS0110/20-3-08IF. Computing resources were provided by CSC-IT Center for Science, Finland.